\documentstyle[12pt]{article}
\def\be{\begin{equation}}
\def\ee{\end{equation}}

\begin{document}

\title{Coherent population trapping in the stochastic limit}

\author{L.Accardi, S.V.Kozyrev}
\maketitle

\begin{abstract}
A 2--level atom with degenerate ground state interacting with a
quantum field is investigated. We show, that the field drives the
state of the atom to a stationary state, which is non--unique, but
depends on the initial state of the system through some conserved
quantities. This non--uniqueness follows from the degeneracy of
the ground state of the atom, and when the ground subspace is
two--dimensional, the family of stationary states will depend on a
one--dimensional parameter. Only one of the stationary states in
this family is a pure state, and this state coincides with the
known population trapped state (zero population in the excited
level $|NC\rangle$). Another one stationary state corresponds to
an equal weight mixture of the excited level $|3\rangle$ and of
the coupled state $|C\rangle$.
\end{abstract}

\section{Introduction}

In the present paper we consider a 2--level atom with a degenerate
ground state (or, equivalently, a 3--level atom with equal
energies of the two lower levels). We prove that the interaction
with radiation drives this atom to a family of stationary states,
depending linearly on a one--dimensional parameter, which varies
in explicitly determined interval. For a particular (extremal)
value of this parameter the stationary state coincides with the
coherent population trapped state, described in
\cite{0}--\cite{3}.

Our starting point are the papers  \cite{1}, \cite{2}, which
discuss coherent population trapping (CPT) in a 3--level
$\Lambda$--system (i.e. a 3-level atom where only two transitions
between the the lower levels 1 and 2 and the higher level 3 are
allowed, and the transition between 1 and 2 is forbidden). CPT is
based on the preparation of atoms in a special coherent
superposition of the two lower states. In \cite{2} it was argued
that the CPT process may be described in the basis of coupled and
non--coupled states defined by \be\label{coupled}
|C\rangle={1\over\sqrt{2}}\left(|1\rangle+|2\rangle\right) \ee
\be\label{noncoupled}
|NC\rangle={1\over\sqrt{2}}\left(|1\rangle-|2\rangle\right) \ee
where the two levels $|1\rangle$, $|2\rangle$, correspond to the
hyperfine split $3S_{1/2}$ sodium ground state, coupled by the
laser fields to a common excited level $|3\rangle$ within
$3P_{1/2}$ levels.

In \cite{1}, \cite{3} it is explained that this scheme, although
highly simplified with respect to the real situation, nevertheless
captures the main physical features of the phenomenon of CPT.

Coherent population trapping (CPT) consists in driving the atomic
state to the superposition wave function $|NC\rangle$, often
called population trapped or dark state cf. \cite{0}--\cite{3}. An
atom driven in this state is transparent, i.e. it does not absorb
the incoming laser radiation, in the sense that transitions from
the trapped state to the excited $|3\rangle$ are forbidden
\cite{Harris}. The idea is that, at resonance and in the
stationary state, all the atomic population is pumped into the
non--coupled state, which is not excited by the laser radiation,
so that the excited state population, hence the emitted
fluorescent intensity reaches the minimum.

In \cite{5} trapped states were used to propose a scheme to
utilize photons for ideal quantum transmission between atoms
located at spatially separated nodes of a quantum network.

Now, the process of driving the atom to the non coupled state is
dynamical and therefore it is natural to try and apply the
stochastic limit technique \cite{book} in order to deduce a (non
phenomenological) master equation for which the $|NC\rangle$ state
is an attractor. The present note essentially confirms this
intuition, but also shows that the situation is more complex than
this. In fact the $|NC\rangle$ state is indeed an attractor for
the reduced dynamics, but not a global one: there exists a one
dimensional parameter family of stationary states and the precise
interval of this parameter is determined in Theorem 4 below. This
means that the set of atomic states is split into nonintersecting
sets, one for each value of the parameter, and each of these sets
is a domain of attraction for the state corresponding to the given
value of the parameter. These stationary states generally are
mixed. Only one of these states is pure and coincides with the
known non--coupled, or population trapped state
(\ref{noncoupled}), investigated in \cite{0}--\cite{3}.

This extends the effect of population trapping and predicts a
dependence of the trapped state on the preparation procedure. This
also extends our ability to control the quantum states of the
atom. In fact, by preparing the initial state, switching the
interaction with the field and (possibly) filtering the excited
state, one can realize a switch between the $|NC\rangle$ and the
$|C\rangle$ states.

In the present note the role of the velocity of the atoms has not
been investigated. It is natural to conjecture, from the analysis
of \cite{1}, \cite{3}, that the selection parameter of the
stationary states depends on this velocity. The explicit form of
this dependence is now under investigation. Also the dependence on
the initial state of the field is investigated. It is shown that
the above scenario can be realized only in equilibrium or non
equilibrium, but non vacuum, states. In the vacuum state new
phenomena, such as quantum beats, may arise (cf. Remark 11 below).

\section{The master equation}

For the investigation of the dynamics of a 3--level
$\Lambda$--system interacting with radiation we use the stochastic
limit approach, \cite{book}. In this approach one introduces a
slow time scale ${t/ \lambda^2}$, where $\lambda$ is a coupling
constant for the interaction of the system with radiation. In the
limit $\lambda\to 0$ the dynamics is given by Langevin and master
equations, cf. \cite{book}, \cite{notes}, which are unambiguously
derived from the original Hamiltonian. For the mathematical
discussion of Langevin and master equations see also \cite{chebo},
\cite{smo}. Evolution of the slow degrees of freedom of the filed
in the stochastic limit approach was considered in \cite{AIK}.

We consider a 3--level system with degenerate (for example,
hyperfine split) ground states $|1\rangle$, $|2\rangle$ and the
excited state $|3\rangle$.

The interaction of the system with the radiation field is
described by the Hamiltonian
\begin{equation}\label{H}
H=H_S+H_R+\lambda H_I
\end{equation}
where the system degrees of freedom are described by the
Hamiltonian $H_S$:
$$
H_S=\varepsilon_1|1\rangle\langle 1|+\varepsilon_1|2\rangle\langle
2|+\varepsilon_3|3\rangle\langle 3|
$$
where $\varepsilon_i$ is the energy of the level $|i\rangle$ (note
that $\varepsilon_1=\varepsilon_2$ ).

The radiation degrees of freedom are described by the Hamiltonian
\begin{equation}\label{e_reser}
H_R=\sum_i\int \omega(k)a_i^*(k)a_i(k) dk
\end{equation}
where $a_i(k)$ is a boson field with a mean zero gauge invariant
Gaussian state characterized by the pair correlations
\begin{equation}\label{e_state}
\langle a_i^*(k)a_j(k')\rangle=N_i(k)\delta_{ij}\delta(k-k')
\end{equation}
and $i$, $j=1,2$ are the polarization indices.

The interaction Hamiltonian $H_I$ is defined as follows
\begin{equation}\label{e_inter}
H_I=\int \sum_{i\alpha}\overline{g_{i\alpha}(k)}
a_i(k)D_{\alpha}^* dk +\hbox{ h.c. }
\end{equation}
where $\alpha$ takes two values 1 and 2 and
$$
D_{1}=|1\rangle\langle3|,\qquad D_{2}=|1\rangle\langle2|
$$

The free evolution of the interaction is equivalent to an
effective free evolution of the boson field of the form
$$
e^{-it(\omega(k)-\omega)}a_i(k)
$$
where $\omega=\varepsilon_3-\varepsilon_1$ is the Bohr frequency,
which is equal to the difference of energies of the two energy
levels.

By the stochastic golden rule \cite{book} the rescaled free
evolution of the field above, in the stochastic limit, becomes a
quantum white noise $b_{i\omega}(t,k)$, or master field satisfying
the commutation relations
\begin{equation}\label{2_5}
[b_{i\omega}(t,k),b^*_{j\omega'}(t',k')]=2\pi\delta_{\omega
,\omega '}\delta_{ij}
\delta(t-t')\delta(\omega(k)-\omega)\delta(k-k')
\end{equation}
and with the mean zero gauge invariant Gaussian state with
correlations:
\begin{equation}\label{cormafi1}
\langle
b^*_{i\omega}(t,k)b_{j\omega'}(t',k')\rangle=2\pi\delta_{\omega
,\omega'}
\delta_{ij}\delta(t-t')\delta(\omega(k)-\omega)\delta(k-k')N_i(k)
\end{equation}
\begin{equation}\label{cormafi2}
\langle
b_{i\omega}(t,k)b^*_{j\omega'}(t',k')\rangle=2\pi\delta_{\omega
,\omega'}
\delta_{ij}\delta(t-t')\delta(\omega(k)-\omega)\delta(k-k')(N_i(k)+1)
\end{equation}

The Schr\"odinger equation becomes a white noise Hamiltonian
equation, cf. \cite{book}, \cite{notes} which when put in normal
order is equivalent to the quantum stochastic differential
equation (QSDE)
\begin{equation}\label{1.21}
dU_t=(-idH(t)-Gdt)U_t \qquad ;\quad t>0
\end{equation}
with initial condition $U_0=1$ and where

(i) $h(t)$ is the white noise Hamiltonian and $dH(t)$, called {\it
the martingale term\/}, is the stochastic differential:
\begin{equation}\label{2_9}
dH(t)=\int^{t+dt}_th(s)ds= \sum_{i\alpha\omega}\left(
D^*_{\alpha}dB_{i\alpha\omega}(t)+
D_{\alpha}dB^*_{i\alpha\omega}(t) \right)
\end{equation}
driven by the quantum Brownian motions
\begin{equation}\label{2_10}
dB_{i\alpha\omega}(t):=\int^{t+dt}_t\int dk\overline
g_{i\alpha}(k)b_{i\omega}(\tau,k)d\tau =:\int^{t+dt}_t
b_{i\omega}(\tau,g_{i\alpha})d\tau
\end{equation}

(ii) The operator $G$, called the {\it drift\/}, is given by
\begin{equation}\label{drift1}
G=\sum_{i\alpha\beta\omega} \left(
(g_{i\alpha}|g_{i\beta})^-_{\omega} D^*_{\alpha}D_{\beta}
+\overline{(g_{i\alpha}|g_{i\beta})}^+_{\omega}
D_{\alpha}D^*_{\beta} \right)
\end{equation}
where the explicit form of the constants
$(g_{i\alpha}|g_{i\beta})^{\pm}_{\omega}$, called the generalized
susceptivities, is:
\begin{equation}\label{1.27a}
(g_{i\alpha}|g_{i\beta})^{-}_{\omega}= -i\int dk\,
\overline{g_{i\alpha}(k)}g_{i\beta}(k)
{N_i(k)+1\over\omega(k)-\omega-i0}
\end{equation}
$$
=\pi\int dk\, \overline{g_{i\alpha}(k)}g_{i\beta}(k) (N_i(k)+1)
\delta(\omega(k)-\omega) -i\,\hbox{P.P.}\,\int dk\,
\overline{g_{i\alpha}(k)}g_{i\beta}(k)
{N_i(k)+1\over\omega(k)-\omega}
$$
\begin{equation}\label{1.27b}
(g_{i\alpha}|g_{i\beta})^{+}_{\omega}=-i\int dk\,
\overline{g_{i\alpha}(k)}g_{i\beta}(k)
{N_i(k)\over\omega(k)-\omega-i0}
\end{equation}
$$
=\pi\int dk\, \overline{g_{i\alpha}(k)}g_{i\beta}(k) N_i(k)
\delta(\omega(k)-\omega) -i\,\hbox{P.P.}\,\int dk\,
\overline{g_{i\alpha}(k)}g_{i\beta}(k)
{N_i(k)\over\omega(k)-\omega}
$$
We will use the notations \be\label{16a}
\hbox{Re}\,(g_{i\alpha}|g_{i\beta})^{+}_{\omega}=\pi\int dk\,
\overline{g_{i\alpha}(k)}g_{i\beta}(k) N_i(k)
\delta(\omega(k)-\omega) \ee \be\label{16b} \hbox{ Im
}\,(g_{i\alpha}|g_{i\beta})^{+}_{\omega}=-i\,\hbox{P.P.}\,\int
dk\, \overline{g_{i\alpha}(k)}g_{i\beta}(k)
{N_i(k)\over\omega(k)-\omega} \ee \be\label{16c}
(g_{i\alpha}|g_{i\beta})^{+}_{\omega}=
\hbox{Re}\,(g_{i\alpha}|g_{i\beta})^{+}_{\omega}+i\hbox{Im}\,(g_{i\alpha}|g_{i\beta})^{+}_{\omega}
\ee \be\label{16d}
\overline{(g_{i\alpha}|g_{i\beta})^{+}_{\omega}}=
\hbox{Re}\,(g_{i\alpha}|g_{i\beta})^{+}_{\omega}-i\hbox{Im}\,(g_{i\alpha}|g_{i\beta})^{+}_{\omega}
\ee

Note that for $\alpha=\beta$ the values (\ref{16a}), (\ref{16b})
coincides with real and imaginary part of the generalized
susceptivities, but the are and not necessarily equal to real and
imaginary parts for $\alpha\ne\beta$. The values in (\ref{16c}),
(\ref{16d}) are related in the following way: the complex
conjugation of $(g_{i\alpha}|g_{i\beta})^{+}_{\omega}$ is equal to
$\overline{(g_{i\beta}|g_{i\alpha})^{+}_{\omega}}$.

We also use the notation \be\label{sumpolarization}
(g_{\alpha}|g_{\beta})^{\pm}_{\omega}=\sum_i(g_{i\alpha}|g_{i\beta})^{\pm}_{\omega}
\ee

\bigskip

\noindent{\bf Remark 1}.\qquad Typically the $g_{i\alpha}$ are
matrix elements (cf. the description in section 4.9.3 of
\cite{book}). Therefore their dependence on the index $\alpha$ is
often unavoidable. Therefore we will develop the theory, as far as
possible, keeping this dependence explicit. In some cases, e.g.
some particular classes of 3--level atoms, the assumption that the
formfactors $g_{i\alpha}$ do not depend on the index $\alpha$, is
justified. In this case the formulae simplify and are easier to
interpret. Thus situation is described in the Section 3 below.

\bigskip

\noindent{\bf Remark 2}.\qquad  Note that if the expectation of
number operators $N_i(k)$ in the reference state depends only on
the dispersion $\omega(k)$ (in this case we will denote this value
$N(\omega)$), then we have the identity \be\label{Romega}
R_{\omega}=
{\hbox{Re}\,(g|g)^+_{\omega}\over\hbox{Re}\,(g|g)^-_{\omega}}={N(\omega)\over
N(\omega)+1}\ee  which shows that this quotient of generalized
susceptivities does not depend on the formfactor $g$ and is a
natural non equilibrium generalization of the Einstein
emission--absorption coefficient (cf. section 5.9 of \cite{book}).

\bigskip

The master equation for the reduced density matrix $\rho(t)$ of
the system for a general discrete system with the dipole
interaction in the stochastic limit approach was found in
\cite{notes}. For the considered degenerate 3--level
$\Lambda$--system it takes the form \be\label{master3}
{d\rho(t)\over dt}= \sum_{j} \biggl(\biggl( i\hbox{ Im
}(g_{j1}|g_{j1})^-_{\omega} [\rho,|3\rangle\langle3|]-i\hbox{ Im
}{(g_{j1}|g_{j1})}^+_{\omega} [\rho,|1\rangle\langle1|]+
$$
$$
+2\hbox{Re}\,(g_{j1}|g_{j1})^-_{\omega}\left(\rho_{33}|1\rangle\langle1|
-{1\over 2} \{\rho,|3\rangle\langle3|\}\right)
+2\hbox{Re}\,{(g_{j1}|g_{j1})}^+_{\omega}
\left(\rho_{11}|3\rangle\langle3| -{1\over 2}
\{\rho,|1\rangle\langle1| \}\right)\biggr)
$$
$$
+ \biggl( i\hbox{ Im }(g_{j2}|g_{j2})^-_{\omega}
[\rho,|3\rangle\langle3|]-i\hbox{ Im }{(g_{j2}|g_{j2})}^+_{\omega}
[\rho,|2\rangle\langle2|]+
$$
$$
+2\hbox{Re}\,(g_{j2}|g_{j2})^-_{\omega}\left(\rho_{33}|2\rangle\langle2|
-{1\over 2} \{\rho,|3\rangle\langle3|\}\right)
+2\hbox{Re}\,{(g_{j2}|g_{j2})}^+_{\omega}
\left(\rho_{22}|3\rangle\langle3| -{1\over 2}
\{\rho,|2\rangle\langle2| \}\right)\biggr)
$$
$$
+\biggl( -i\hbox{ Im }{(g_{j1}|g_{j2})}^+_{\omega}
[\rho,|1\rangle\langle2|]+
2\hbox{Re}\,(g_{j1}|g_{j2})^-_{\omega}\rho_{33}|2\rangle\langle1|
+2\hbox{Re}\,{(g_{j1}|g_{j2})}^+_{\omega}
\left(\rho_{21}|3\rangle\langle3| -{1\over 2}
\{\rho,|1\rangle\langle2| \}\right)\biggr)
$$
$$
+\biggl( -i\hbox{ Im }{(g_{j2}|g_{j1})}^+_{\omega}
[\rho,|2\rangle\langle1|]+
2\hbox{Re}\,(g_{j2}|g_{j1})^-_{\omega}\rho_{33}|1\rangle\langle2|
+2\hbox{Re}\,{(g_{j2}|g_{j1})}^+_{\omega}
\left(\rho_{12}|3\rangle\langle3| -{1\over 2}
\{\rho,|2\rangle\langle1| \}\right)\biggr)\biggr) \ee where, as
usual $[a,b]=ab-ba$ and $\{a,b\}=ab+ba$.

One of our main results is the following separation of the density
matrix into parts corresponding to invariant subspaces of the
evolution.

\bigskip

\noindent{\bf Lemma 1}.\qquad{\sl The vector space $H(3)$ of the
Hermitian $3\times 3$ matrices is the direct sum of two subspaces,
$V_0$, $V_1$, which are invariant under the evolution, defined by
(\ref{master3}):
$$
H(3)=V_0\oplus V_1
$$

A linear basis of $V_0$ is given by $\{|2\rangle\langle3|$,
$|3\rangle\langle2|$, $|3\rangle\langle1|$,
$|1\rangle\langle3|\}$. Any matrix in this space decays
exponentially to zero under the reduced evolution if the real
parts of generalized susceptivities (\ref{16a}) (for the indices
$\alpha=1,2$ and $\beta=3$ and vice versa) are non zero.

A linear basis of $V_1$ is given by $\{|2\rangle\langle1|$,
$|1\rangle\langle2|$, $|3\rangle\langle3|$, $|1\rangle\langle1|$,
$|2\rangle\langle2|\}$. This space contains all the stationary
states for the evolution. }

\bigskip

\noindent{\it Proof}\qquad Direct verification from the right hand
side of ({master3}).

\bigskip

\noindent{\bf Remark 3}.\qquad Notice that the space $V_1={\bf
C}|3\rangle\langle 3|\oplus M$, where $M$ is the $2\times 2$
matrix algebra generated by $|1\rangle\langle 2|$, is itself a
$*$--algebra.

\bigskip

\noindent{\bf Remark 4}.\qquad From Lemma 1 we deduce that the
evolution of the density matrix $\rho(t)$ can be split into the
sum of two evolutions $\rho_0(t)$ and $\rho_1(t)$, where
$\rho_0(t)$ is an off diagonal matrix and $\rho_1(t)$ is a density
matrix.
$$
\rho(t)= \rho_0(t)+\rho_1(t)=
         \pmatrix{0     & \rho_{32}(t) & \rho_{33}(t)\cr
                  \rho_{23}(t)     & 0 & 0 \cr
                  \rho_{13}(t)     & 0 & 0 \cr}+
\pmatrix{\rho_{33}(t) & 0            & 0\cr
                  0             & \rho_{22}(t)& \rho_{21}(t)\cr
                  0             & \rho_{12}(t)& \rho_{11}(t)\cr}
$$
Moreover, $\|\rho_0(t)\|\le e^{-ct}$, where $2c=\hbox{ min
}\hbox{Re}\,(g_{j\alpha}|g_{j\alpha})^{\pm}_{\omega}$, $j=1,2$,
$\alpha=1,2$, i.e. the off--diagonal part of $\rho(t)$ (in $V_0$)
decays exponentially whenever $c>0$.

\section{Stationary states for the master equation}

In the present section we will describe the set of stationary
states for the evolution, generated by the master equation
(\ref{master3}). By lemma 1 the invariant states of
(\ref{master3}) belong to space $V_1$. On this subspace equation
(\ref{master3}) reduces to the following system of three
differential equations \be\label{master22} {d\rho_{22}(t)\over
dt}= 2\hbox{Re}\,(g_{2}|g_{2})^-_{\omega}\rho_{33}
-2\hbox{Re}\,{(g_{2}|g_{2})}^+_{\omega} \rho_{22}
-{(g_{1}|g_{2})}^+_{\omega} \rho_{21}
-\overline{(g_{1}|g_{2})^+_{\omega}} \rho_{12} \ee
\be\label{master11}  {d\rho_{11}(t)\over dt}=
2\hbox{Re}\,(g_{1}|g_{1})^-_{\omega}\rho_{33}
-2\hbox{Re}\,{(g_{1}|g_{1})}^+_{\omega} \rho_{11}
-\overline{(g_{2}|g_{1})^+_{\omega} }\rho_{21}
-{(g_{2}|g_{1})}^+_{\omega} \rho_{12} \ee \be\label{master12}
{d\rho_{12}(t)\over dt}=
-\left(\overline{(g_{1}|g_{1})^+_{\omega}}
+{(g_{2}|g_{2})}^+_{\omega}\right) \rho_{12}
-{(g_{1}|g_{2})}^+_{\omega}
\rho_{11}-\overline{(g_{2}|g_{1})^+_{\omega}}\rho_{22}
 +2\hbox{Re}\,(g_{2}|g_{1})^-_{\omega}\rho_{33}
\ee which together with the normalization condition
$$
\rho_{11}+\rho_{22}+\rho_{33}=1
$$
the conjugation rule
$$
\rho_{12}^*=\rho_{21},\quad \rho_{11},\rho_{22},\rho_{33}\in {\bf
R}
$$
and the conditions of positivity of the density matrix discussed
in the following Lemma, form the set of equations determining the
evolution of density matrix.

\bigskip

\noindent{\bf Lemma 2}.\qquad {\sl The Hermitian matrix
$$
\rho=\pmatrix{\rho_{33} & 0            & 0\cr
                  0             & \rho_{22}& \rho_{21}\cr
                  0             & \rho_{12}& \rho_{11}\cr}
$$
is a density matrix iff the diagonal elements satisfy
\be\label{28a} \rho_{11}+\rho_{22}+\rho_{33}=1,\qquad \rho_{11},
\rho_{22}, \rho_{33}\ge 0 \ee and the off-diagonal elements
satisfy \be\label{28b} \rho_{12}^*=\rho_{21}, \qquad
|\rho_{12}|^2\le \rho_{11}\rho_{22} \ee

}

\bigskip

\noindent{\bf Lemma 3}.\qquad {\sl When the susceptivities
${(g_{\alpha}|g_{\beta})}^\pm_{\omega}$ do not depend on $\alpha$,
$\beta$ (in this case we denote them ${(g|g)}^\pm_{\omega}$) the
system (\ref{master22})--(\ref{master12}) of linear equations
determining the evolution of the atom has the conservation law
\be\label{ConsLaw}
\rho_{11}(t)+\rho_{22}(t)=\rho_{12}(t)+\rho_{21}(t)+C;\qquad
\forall t \ee Moreover, if $\hbox{Re} (g|g)^-_{\omega}> 0$, then
$\rho_{12}(t)+\rho_{21}(t)$ converges exponentially in time to the
stationary value \be\label{24b}
\rho_{12}+\rho_{21}={4\hbox{Re}\,(g|g)^-_{\omega}(1-C)
-2\hbox{Re}\,{(g|g)}^+_{\omega} C \over
4(\hbox{Re}\,(g|g)^-_{\omega}+\hbox{Re}\,{(g|g)}^+_{\omega})}
={1-C-CR_{\omega}/2\over 1+R_{\omega}}\ee where $C$ is a real
constant. }

\bigskip

\noindent{\it Proof}\qquad The conservation law (\ref{ConsLaw})
follows from the identity
$$
{d\rho_{11}(t)\over dt}+{d\rho_{22}(t)\over
dt}={d\rho_{12}(t)\over dt}+{d\rho_{21}(t)\over dt}
$$
It implies that for fixed $C$ the evolution of the system is
characterized by the real function of time
$\rho_{12}(t)+\rho_{21}(t)$ which we denote $2s(t)$:
$$
s(t)={1\over 2}\left(\rho_{12}(t)+\rho_{21}(t)\right)
$$
With this notation the system (\ref{master22})--(\ref{master12})
implies that \be\label{decay} {d s(t)\over dt}
=-4\left(\hbox{Re}\,(g|g)^-_{\omega}+\hbox{Re}\,{(g|g)}^+_{\omega}\right)s(t)+2\hbox{Re}\,(g|g)^-_{\omega}(1-C)
-\hbox{Re}\,{(g|g)}^+_{\omega} C \ee If $\hbox{Re}
(g|g)^-_{\omega}> 0$, then equation (\ref{decay}) implies the
exponential decay of $s(t)$ to the stationary value (\ref{24b})
and this proves the lemma.

\bigskip

\noindent{\bf Remark 5}.\qquad Note that the condition $\hbox{Re}
(g|g)^-_{\omega}> 0$ means that
$$
\int \overline{g(k)}g(k) \delta(\omega(k)-\omega) dk\ne 0
$$
which is automatically satisfied when the support of the
formfactor $g(k)$ intersects with the resonant surface
$\omega(k)=\omega$.

\bigskip

The stationary solution of the system
(\ref{master22})--(\ref{master12}) is determined by the system of
equations \be\label{master22s}
2\hbox{Re}\,(g_{2}|g_{2})^-_{\omega}(\rho_{11}+\rho_{22})
+2\hbox{Re}\,{(g_{2}|g_{2})}^+_{\omega} \rho_{22}
+{(g_{1}|g_{2})}^+_{\omega} \rho_{12}^*
+\overline{(g_{1}|g_{2})^+_{\omega}
}\rho_{12}=2\hbox{Re}\,(g_{2}|g_{2})^-_{\omega}\ee
\be\label{master11s}
2\hbox{Re}\,(g_{1}|g_{1})^-_{\omega}(\rho_{11}+\rho_{22})
+2\hbox{Re}\,{(g_{1}|g_{1})}^+_{\omega}\rho_{11}
+\overline{(g_{2}|g_{1})^+_{\omega}} \rho_{12}^*
+{(g_{2}|g_{1})}^+_{\omega}
\rho_{12}=2\hbox{Re}\,(g_{1}|g_{1})^-_{\omega}\ee
\be\label{master12s}\left(\overline{(g_{1}|g_{1})^+_{\omega}}
+{(g_{2}|g_{2})}^+_{\omega}\right)\rho_{12}=
{-{(g_{1}|g_{2})}^+_{\omega}
\rho_{11}-\overline{(g_{2}|g_{1})^+_{\omega}}\rho_{22}
 +2\hbox{Re}\,(g_{2}|g_{1})^-_{\omega}\rho_{33}} \ee

\bigskip

\noindent{\bf Remark 6}.\qquad  For different formfactors
$g_{\alpha}(k)$ the system (\ref{master22})--(\ref{master12}) may
have different behaviors. In the generic case for $g_1\ne g_2$ the
stationary solution is unique. For instance when $g_1$ is
orthogonal to $g_2$ (in the sense of the bilinear form
$(g_{1}|g_{2})^+_{\omega}$), then the determinant of the system
(\ref{master22s}), (\ref{master11s}) reduces to
$$
-\left(
2\hbox{Re}\,(g_{2}|g_{2})^+_{\omega}2\hbox{Re}\,(g_{1}|g_{1})^-_{\omega}+
2\hbox{Re}\,(g_{2}|g_{2})^-_{\omega}2\hbox{Re}\,(g_{1}|g_{1})^+_{\omega}+
2\hbox{Re}\,(g_{2}|g_{2})^+_{\omega}2\hbox{Re}\,(g_{1}|g_{1})^+_{\omega}\right)
$$
and whenever this determinant is non-zero, the solution is unique.

\bigskip

When $g_1= g_2$ the solution is non--unique due to Lemma 3.

Now we are ready to formulate the following theorem describing the
structure of the stationary density matrices.

\bigskip

\noindent{\bf Remark 7}.\qquad If for $\alpha,\beta=1,2$
$$
(g_{\alpha}|g_{\beta})^+_{\omega}=0
$$
in particular, in the Fock case, the stationary solutions of
(\ref{master22})--(\ref{master12}) (neglecting the trivial case
when also $(g_{\alpha}|g_{\beta})^-_{\omega}=0$) is characterized
by the single condition
$$
\rho_{11}+\rho_{22}=1,\quad \rho_{11},\rho_{22}\ge 0
$$
so that $\rho_{33}=0$ and $\rho_{12}$ is arbitrary and subject
only to the constraints (\ref{28b}).

\bigskip

\noindent{\bf Theorem 4}.\qquad {\sl For
${(g_{\alpha}|g_{\beta})}^\pm_{\omega}$ not depending on $\alpha$,
$\beta$ and when \be\label{gt0} \hbox{Re}\,(g|g)^{+}_{\omega}>0
\ee the system (\ref{master22s})--(\ref{master12s}) of linear
equations determining the stationary state of the atom possesses a
family of solutions parameterized by the one--dimensional
parameter: \be\label{densitymatrix} \rho=\pmatrix{\rho_{e} & 0 &
0\cr
                  0             & \rho_{g}& s\cr
                  0             & s& \rho_{g}\cr} \ee
where, in the notations (\ref{Romega}) \be\label{element1}
\rho_{e}={2\hbox{Re}\,(g|g)^+_{\omega} (1+2s)\over
4\hbox{Re}\,(g|g)^-_{\omega}+2\hbox{Re}\,(g|g)^+_{\omega}}={(1+2s)R_{\omega}\over
2+ R_{\omega} }\ee \be\label{element2}
\rho_{g}={2\hbox{Re}\,(g|g)^-_{\omega}-2\hbox{Re}\,(g|g)^+_{\omega}
s\over
4\hbox{Re}\,(g|g)^-_{\omega}+2\hbox{Re}\,(g|g)^+_{\omega}}={1-sR_{\omega}\over
2+ R_{\omega} } \ee The admissible values of the parameter $s$ are
precisely those for which \be\label{interval} {1\over 2(1+
R_{\omega}) }={1\over 2}\left(
1+{\hbox{Re}\,(g|g)^+_{\omega}\over\hbox{Re}\,(g|g)^-_{\omega}}\right)^{-1}\ge
s\ge -{1\over 2} \ee

Moreover, if (\ref{gt0}) is satisfied, the solution of the system
(\ref{master22})--(\ref{master12}) converges, as $t\to\infty$, to
the stationary state (\ref{densitymatrix}). }

\bigskip

\noindent{\it Proof}\qquad If $g_1=g_2=g$, then
(\ref{master22s})--(\ref{master12s}) take respectively the form:
\be\label{m22} 2\hbox{Re}\,(g|g)^-_{\omega}(\rho_{11}+\rho_{22})
+2\hbox{Re}\,{(g|g)}^+_{\omega} \rho_{22} +{(g|g)}^+_{\omega}
\rho_{21} +\overline{(g|g)^+_{\omega}
}\rho_{12}=2\hbox{Re}\,(g|g)^-_{\omega}\ee \be\label{m11}
2\hbox{Re}\,(g|g)^-_{\omega}(\rho_{11}+\rho_{22})
+2\hbox{Re}\,{(g|g)}^+_{\omega}\rho_{11}
+\overline{(g|g)^+_{\omega}} \rho_{21} +{(g|g)}^+_{\omega}
\rho_{12}=2\hbox{Re}\,(g|g)^-_{\omega}\ee
\be\label{m12}2\hbox{Re}\,{(g|g)}^+_{\omega}\rho_{12}=
{-{(g|g)}^+_{\omega}
\rho_{11}-\overline{(g|g)^+_{\omega}}\rho_{22}
 +2\hbox{Re}\,(g|g)^-_{\omega}\rho_{33}} \ee
Taking the differences of (\ref{m22}), (\ref{m11}) and of
(\ref{m12}) and its conjugate, we obtain
$$
2\left(\hbox{Re}\,{(g|g)}^+_{\omega}\right)
(\rho_{22}-\rho_{11})+2i\left(\hbox{Im}\,{(g|g)}^+_{\omega}\right)
(\rho_{21}-\rho_{12})=0
$$
$$
2\left(\hbox{Re}\,{(g|g)}^+_{\omega}\right)
(\rho_{21}-\rho_{12})+2i\left(\hbox{Im}\,{(g|g)}^+_{\omega}\right)
(\rho_{22}-\rho_{11})=0
$$
In the following we will not indicates the brackets at
$\hbox{Re}(g|g)$ and $\hbox{Im}(g|g)$.

Taking the sum of two equations above and dividing by two, we get
$$
{(g|g)}^+_{\omega}(\rho_{22}-\rho_{11}+\rho_{21}-\rho_{12})=0
$$
If ${(g|g)}^+_{\omega}\ne 0$, then since $\rho_{22}-\rho_{11}$ is
real, and $\rho_{12}-\rho_{21}$ is imaginary, we obtain
\be\label{equal} \rho_{22}=\rho_{11},\qquad \rho_{21}=\rho_{12}
\ee Then, the sum of (\ref{m22}) and (\ref{m11}) takes the form
\be\label{sumof}
2\hbox{Re}\,{(g|g)}^+_{\omega}(\rho_{11}+\rho_{22}+\rho_{12}+\rho_{21})=4\hbox{Re}\,{(g|g)}^-_{\omega}\rho_{33}
\ee Equations (\ref{equal}) and (\ref{sumof}) imply that any
stationary density matrix must satisfy the following condition:
\be\label{ex2} 2\hbox{Re}\,{(g|g)}^+_{\omega}
(\rho_{11}+\rho_{12})= 2\hbox{Re}\,(g|g)^-_{\omega}\rho_{33} \ee
Since under the condition (\ref{equal}) the equations (\ref{m22})
and (\ref{m11}) coincide, equations (\ref{equal}), (\ref{ex2})
describe the general stationary solution for
(\ref{master22})--(\ref{master12}). Using (\ref{equal}),
(\ref{ex2}) and (\ref{28a}), we obtain \be\label{answer}
\rho_{11}=\rho_{22}={2\hbox{Re}\,(g|g)^-_{\omega}-2\hbox{Re}\,{(g|g)}^+_{\omega}\rho_{12}\over
4\hbox{Re}\,(g|g)^-_{\omega}+2\hbox{Re}\,{(g|g)}^+_{\omega}},\qquad
\rho_{33}={2\hbox{Re}\,{(g|g)}^+_{\omega} +
4\hbox{Re}\,{(g|g)}^+_{\omega}\rho_{12}\over
4\hbox{Re}\,(g|g)^-_{\omega}+2\hbox{Re}\,{(g|g)}^+_{\omega}} \ee
In particular, $\rho_{12}$ must be a real number. From
(\ref{answer}) and (\ref{28b}) one sees that the positivity of the
density matrix is equivalent to inequalities \be\label{positivity}
{1\over 2}\left(
1+{\hbox{Re}\,(g|g)^+_{\omega}\over\hbox{Re}\,(g|g)^-_{\omega}}\right)^{-1}\ge
\rho_{12}\ge -{1\over 2} \ee Conversely, taking any real value of
$\rho_{12}$ satisfying (\ref{positivity}) and determining
$\rho_{11}$ and $\rho_{33}$ by (\ref{answer}), one obtains a
stationary state for the master equation
(\ref{master22})--(\ref{master12}).

Let us now prove the convergence of the system to a stationary
state. The system (\ref{master22})--(\ref{master12}) implies
\be\label{31c} {d\over
dt}(\rho_{22}-\rho_{11})=-2\hbox{Re}\,{(g|g)}^+_{\omega}
(\rho_{22}-\rho_{11})+2i\hbox{ Im
}{(g|g)}^+_{\omega}(\rho_{12}-\rho_{21}) \ee \be\label{31d}
{d\over dt}(\rho_{12}-\rho_{21})=
-2\hbox{Re}\,{(g|g)}^+_{\omega}(\rho_{12}-\rho_{21}) +2i\hbox{ Im
}{(g|g)}^+_{\omega} (\rho_{22}-\rho_{11}) \ee Adding these two
equations we see that \be\label{31b}
\rho_{22}-\rho_{11}+\rho_{12}-\rho_{21}=\hbox{ const
}e^{t\left(-2\hbox{Re}\,{(g|g)}^+_{\omega}+2i\hbox{ Im
}{(g|g)}^+_{\omega}\right)} \ee For the case
$\hbox{Re}\,{(g|g)}^+_{\omega}>0$ the linear combination
(\ref{31b}) converges exponentially to zero.  Since
$\rho_{22}-\rho_{11}$ is real and $\rho_{12}-\rho_{21}$ is
imaginary, we obtain that (\ref{31b}) converges to the state where
$\rho_{22}=\rho_{11}$ and $\rho_{12}=\rho_{21}$ (and therefore
real).

Then, applying Lemma 3, we get that $\rho_{12}=\rho_{21}$,
$\rho_{22}=\rho_{11}$ and $\rho_{33}$ converge to stationary
values, which are controlled by the stationary value $s={1\over
2}(\rho_{12}+\rho_{21})$.

This finishes the proof of the theorem.

\bigskip

\noindent{\bf Remark 8}.\qquad Note that if
$\hbox{Re}\,(g|g)^{+}_{\omega}=0$ and
$\hbox{Im}\,(g|g)^{+}_{\omega}\ne 0$ then (\ref{31b}) implies that
the system does not converge to a stationary state but has an
oscillatory behavior.

\bigskip

Since the generalized susceptivities are given by the expression
$$
\hbox{Re}\,(g_{i}|g_{i})^{+}_{\omega}=\pi\int  |g_{i}(k)|^2 N_i(k)
\delta(\omega(k)-\omega)dk
$$
$$
\hbox{Re}\,(g_{i}|g_{i})^{-}_{\omega}=\pi\int  |g_{i}(k)|^2
(N_i(k)+1) \delta(\omega(k)-\omega)dk
$$
$$
\hbox{Re}\,(g|g)^{\pm}_{\omega}=\sum_{i}\hbox{Re}\,(g_{i}|g_{i})^{\pm}_{\omega}
$$

It follows that one has inequality
$$
\hbox{Re}\,{(g|g)}^-_{\omega}> \hbox{Re}\,{(g|g)}^+_{\omega}
$$

One can see that for high intensity of radiation, i.e. when
$N_i(k)>>1$, one can put ${N_i(k)\over N_i(k)+1}=1$. In this case
the solution (\ref{densitymatrix}), (\ref{interval}) will be
simplified as follows
$$
\rho={}\pmatrix{{ 1+2s\over 3} & 0 & 0\cr
                  0             & {1- s\over
3}& s\cr
                  0             & s& {1- s\over
3}\cr},\qquad {1\over 4}\ge s\ge -{1\over 2}
$$

The most interesting states correspond to the extremal values of
the parameter $\rho_{12}$. The minimal value of $\rho_{12}$ is
$-{1\over 2}$, which correspond to the density matrix for the pure
state $|NC\rangle$:
$$
\rho_{\hbox{min}}={1\over 2} \pmatrix{0 & 0 & 0\cr
         0 & 1 &-1\cr
         0 &-1 & 1\cr}=|NC\rangle\langle NC|={1\over
2}\left(|1\rangle\langle1|+|2\rangle\langle2|-|1\rangle\langle2|-|2\rangle\langle1|\right)
$$
where the vector
$$
|NC\rangle={1\over\sqrt{2}}\left(|1\rangle-|2\rangle\right)
$$
is exactly the coherent population trapped state
(\ref{noncoupled}), discussed in the literature
\cite{0}--\cite{2}. In the same approximation the maximal value
$\rho_{12}={1\over 4}$ corresponds to the density matrix
$$
\rho_{\hbox{max}}={1\over 4} \pmatrix{2 & 0 & 0\cr
         0 & 1 & 1\cr
         0 & 1 & 1\cr}={1\over
2}|3\rangle\langle 3|+{1\over 2}|C\rangle\langle C|
$$
This state is mixed, but the state of the reduced system
corresponding to levels $|1\rangle$ and $|2\rangle$ is pure with
the state vector
$$
|C\rangle={1\over\sqrt{2}}\left(|1\rangle+|2\rangle\right)
$$
which coincides with the coupled state (\ref{coupled}).

Thus the application of the stochastic limit approach allows to
generalize the coherent population trapping phenomenon. We see
that the family of stationary density matrices realizes a
continuous interpolation between the coupled and the non--coupled
state.

\bigskip

\noindent{\bf Remark 9}.\qquad  To distinguish experimentally
different population trapped stationary states, one can measure
the following observable
$$
A=|1\rangle\langle 2|+|2\rangle\langle 1|
$$
which for instance may describe the interaction of the hyperfine
split levels with a magnetic field.

In fact the measurement of $A$ in the generic stationary state
gives
$$
\hbox{ tr }\rho A=\rho_{12}+\rho_{21}=2s
$$
and the different trapped states give different mean values of
$A$.

\bigskip

\noindent{\bf Remark 10}.\qquad Equations (\ref{element2}),
(\ref{interval}) imply that the minimum ground state population is
achieved when the parameter $s$ is maximum, i.e.
$$
\rho^{min}_{g}={1\over 2+R_{\omega}}\left(1-{R_{\omega}\over
2+2R_{\omega}}\right)={1\over 2}{1\over 1+R_{\omega}}={1\over
2}{N({\omega})+1\over 2N({\omega})+1}>{1\over 4}
$$
Since the ground level population is $2\rho_g$, it follows that in
any stationary state at least $1/2$ of the population is in the
ground level. On the other hand the above chain of identities
shows, in that the region of high radiation intensity
$N(\omega)>>1$, the estimate $\rho^{min}_{g}={1\over 4}$ is almost
exact and therefore we can conclude that, in this region, for any
stationary state, the population of the excited state is about
${1/ 2}$.

\bigskip

\noindent{\bf Remark 11}.\qquad Consider the regime when there is
no decay to the stationary state and we have the oscillations.
This regime is possible when
$\hbox{Re}\,(g_{j\alpha}|g_{j\beta})^+_{\omega}=0$. We consider
again the case when the susceptivities
${(g_{\alpha}|g_{\beta})}^\pm_{\omega}$ do not depend on $\alpha$,
$\beta$ and all
$\hbox{Re}\,(g_{j\alpha}|g_{j\beta})^+_{\omega}=0$.

By Remark 5 after Lemma 3 it is natural to assume that
$\hbox{Re}\,(g|g)^-_{\omega}>0$ and there is a convergence of
$s(t)={1\over 2}\left(\rho_{12}(t)+\rho_{21}(t)\right)$ to its
stationary value $-{1\over 2}\le s\le {1\over 2}$. Analyzing the
system equations for the density matrix, one can check that in the
considered case the dynamics in the invariant subspace $V_1$ is
described by equation (\ref{31b}), which takes the form
\be\label{oscillate}
\rho_{22}-\rho_{11}+\rho_{12}-\rho_{21}=\hbox{ const }e^{2it\hbox{
\tiny Im }{(g|g)}^+_{\omega}} \ee where $\rho_{ij}$ are complex
numbers satisfying Lemma 2.

This kind of pure oscillatory behavior without damping is related
to the quantum beating.

When $\hbox{Re}\,(g|g)^-_{\omega}>0$, the off--diagonal matrix
elements $\rho_{13}$, $\rho_{23}$ decay exponentially by Remark 4,
cf. \cite{notes}. We see that in the regime
$\hbox{Re}\,(g|g)^-_{\omega}>0$, $\hbox{Re}\,(g|g)^+_{\omega}=0$
(which is satisfied, for instance in the Fock (vacuum) state) the
behavior of the 3--level degenerate $\Lambda$--system for large
times is described by the oscillations (\ref{oscillate}), when
$\rho_{13}=\rho_{23}=\rho_{31}=\rho_{32}=0$ and $s(t)={1\over
2}\left(\rho_{12}(t)+\rho_{21}(t)\right)=\hbox{ const}$, $-{1\over
2}\le s\le {1\over 2}$.

\bigskip\bigskip

In conclusion: in the present paper we investigated the
interaction of an atom with a degenerate ground state with a
quantum field. We find (under natural conditions for the
formfactors), that the evolution drives the atom exponentially to
a stationary state. This stationary state is not unique, and the
family of stationary states may be parameterized by a
one--dimensional parameter. For a special (minimal) value of this
parameter the obtained stationary state is pure and coincides with
the population trapped state, known in the literature
\cite{0}--\cite{2}. The obtained results show the possibility of
emergence of mixed stationary states, which continuously
interpolate between the coupled and the non--coupled states. This
difference can be experimentally detected.

In the case of special states (the Fock state) also the
oscillatory behavior (\ref{oscillate}) is possible.

\bigskip

\centerline{\bf Acknowledgements}

Sergei Kozyrev is grateful to Centro Vito Volterra and Luigi
Accardi for kind hospitality. The authors are grateful to Kentaro
Imafuku for stimulating discussions. This work has been partly
supported by INTAS YSF 2002--160 F2, CRDF (grant
UM1--2421--KV--02), and The Russian Foundation for Basic Research
(projects 02--01--01084 and 00--15--97392).

\end{document}